\documentclass[twocolumn]{aastex631}
\usepackage{amsmath}
\usepackage{amssymb}
\usepackage{hyperref}
\usepackage{soul}
\usepackage{graphicx}

\begin{document}
\title{Constraining the jet composition of GRB 221009A with the prompt TeV emission limit}

\author[0000-0002-0170-0741]{Cui-Yuan Dai}
\affiliation{School of Astronomy and Space Science, Nanjing University, Nanjing 210093, China; xywang@nju.edu.cn; ryliu@nju.edu.cn}
\affiliation{Key Laboratory of Modern Astronomy and Astrophysics (Nanjing University), Ministry of Education, Nanjing
210093, China}

\author[0000-0002-5881-335X]{Xiang-Yu Wang}
\affiliation{School of Astronomy and Space Science, Nanjing University, Nanjing 210093, China; xywang@nju.edu.cn; ryliu@nju.edu.cn}
\affiliation{Key Laboratory of Modern Astronomy and Astrophysics (Nanjing University), Ministry of Education, Nanjing
210093, China}

\author[0000-0003-1576-0961]{Ruo-Yu Liu}
\affiliation{School of Astronomy and Space Science, Nanjing University, Nanjing 210093, China; xywang@nju.edu.cn; ryliu@nju.edu.cn}
\affiliation{Key Laboratory of Modern Astronomy and Astrophysics (Nanjing University), Ministry of Education, Nanjing
210093, China}

\author[0000-0002-9725-2524]{Bing Zhang}
\affiliation{Nevada Center for Astrophysics, University of Nevada, Las Vegas, NV 89154, USA; bing.zhang@unlv.edu}
\affiliation{Department of Physics and Astronomy, University of Nevada Las Vegas, Las Vegas, NV 89154, USA}

\begin{abstract}
Recent LHAASO observations of the prompt emission phase of the brightest-of-all-time GRB 221009A  imposes a stringent limit on the flux ratio between the TeV and MeV emissions, $F_{\rm TeV}/F_{\rm MeV}\le   2\times10^{-5}$, during the period $220 \operatorname{-}230\, {\rm s}$ after the trigger. bf This period covers the peak of the main MeV burst and is just before the TeV afterglow emerges. Within the framework of internal shocks, we study the internal $\gamma\gamma$ absorption in GRB 221009A by generating a set of synthetic bursts in a simulation that reproduces the observed feature of GRB 221009A. We find that the $\gamma\gamma$ absorption does not lead to an exponential cutoff, but rather a power-law spectrum, consistent with previous works. We further find that the attenuation due to $\gamma\gamma$ absorption alone cannot explain the flux limit ratio of GRB 221009A, suggesting a low ratio between synchrotron self-Compton (SSC) and synchrotron emission outputs. This requires the magnetic field energy density to be much larger than the synchrotron photon energy density so that the SSC flux is greatly suppressed. This indicates that the jet composition of GRB 221009A is likely Poynting-flux-dominated.

\end{abstract}

\section{Introduction}
The traditional “fireball” model of gamma-ray bursts (GRBs) assumes that a fireball is formed due to the initial energy release in a catastrophic event (see \cite{Kumar2015} for a review). Then the fireball expands under its own thermal pressure and gets accelerated to a relativistic speed. Most of the thermal energy is converted to the kinetic energy
of the outflow. 
A fraction of the kinetic energy is then dissipated in the so-called “internal shocks”, where the internal collisions among the GRB ejecta produce shocks, leading to particle acceleration and producing the prompt GRB.  An alternative model is that the GRB ejecta carries a dynamically important magnetic field
component \citep{Meszaros1997, Lyutikov2003, Narayan2009,zhang2011}, i.e., $\sigma>1$, where $\sigma$ is the ratio between the Poynting flux and
matter (baryonic) flux.  In this model, the GRB radiation is thought to be powered by the
dissipation of the magnetic field energy in the ejecta.

The most popular emission mechanism of the prompt emission is synchrotron radiation from relativistic electrons accelerated in internal shocks or magnetic dissipation \citep{Meszaros1994,Uhm2014,ZhangBing2014}, which radiate in the magnetic field within the shocked plasma. 
In the optically-thin synchrotron scenario for the prompt emission, the SSC emission produced by the same population of relativistic electrons is naturally expected to generate GeV to TeV gamma-rays \citep{Guetta2003, Gupta2007}.
The intensity of the inverse Compton component depends on the intensity of the magnetic field \citep{Bosnjak2009},   thus the SSC component can be used to diagnose the magnetic field strength and hence the composition of the GRB jet. 

On the other hand, the observed flux of high-energy emission from the SSC component is also sensitive to the  $\gamma\gamma$ opacity, as high-energy gamma-rays suffer from pair production absorption with low-energy photons in the prompt emission.  It is usually thought that the internal absorption leads to an exponential drop in the flux at the high-energy end, and thus a moderate optical depth would largely suppress the high-energy emission flux. 
However,  \cite{Granot2008} argued that, due to the time evolution of the $\gamma\gamma$ opacity, the spectral shape of the $\gamma\gamma$ cutoff in the time-integrated spectrum is not exponential, but rather power-law like. \cite{Aoi2010} studied the time-integrated spectrum from internal shocks
with a numerical approach and found that a broken power law spectrum is formed in practical observations that integrate emissions from
different internal shocks.  \cite{Hascoet} considered a more realistic
calculation of the $\gamma\gamma$ opacity, which takes into account the time, space and direction dependent
photon field existing in the outflow, and find that the $\gamma\gamma$ opacity shows strong variations for a multiple-pulse GRB. Consequently, the $\gamma\gamma$ cutoff is much closer to a power-law steepening. In this case, the suppression due to  $\gamma\gamma$ opacity may not be extreme, and it may be still possible to observe the SSC component, especially for bright GRBs.

The brightest-of-all-time GRB 221009A serendipitously occurred within the field of view of the Large High Altitude Air Shower Observatory (LHAASO) \citep{LHAASO2023}.  GRB 221009A was observed by LHAASO during the main burst phase, yielding a differential flux limit of $\sim 6\times 10^{-8}\,{\rm erg\,cm^{-2}\,s^{-1}}$ (after the EBL correction) at $\sim 1$ TeV from $T_0+220\,\textrm{s}$ to $T_0+230\,\textrm{s}$, where $T_0$ is the trigger time \citep{LHAASO2023}.   This period covers the rising part and the peak of the main MeV burst, and is just before the TeV afterglow emerges. This is the period when the constraint on the TeV/MeV ratio in the prompt phase is the most severe.    Compared to the averaged MeV flux during the same period,  the flux ratio is $\bar{R}\equiv F_{\rm TeV}/F_{\rm MeV}\le   2\times10^{-5}$ \citep{LHAASO2023}. {This represents the strongest constraint on the ratio of TeV flux to MeV flux in GRBs during the prompt emission.  Therefore, it provides a unique opportunity to study the SSC emission and constrain the magnetic field in the prompt emission phase.}

In this paper, we consider the simplest internal shock scenario to study this problem. By constraining the magnetic field strength with the prompt TeV emission limit in the internal shock model, we test the validity of this model. The paper is organized as follows. In Sect.\ref{sec_internal_shock_modeling}, through the simulation of the internal shock,  we generate a synthetic burst that can reproduce the observed feature of GRB 221009A. Then we study the $\gamma\gamma$ opacity in GRB 221009A in Sect.\ref{sec_opacity}. In Sect.\ref{sec_constrain}, by comparing the SSC emission after the $\gamma\gamma$ absorption with the TeV flux limit of GRB 221009A measured by LHAASO, we put constraints on the intensity of the magnetic field in the emitting region of the prompt emission. In Sect.\ref{mangetic dissipation}, we briefly discuss the implication of the TeV limit for the magnetic dissipation model. Finally, we give discussions and conclusions in Sect.\ref{summary}.

\section{Internal shock simulation of GRB 221009A} \label{sec_internal_shock_modeling}
We model the internal shock process following the approach in \cite{Kobayashi1997}. We assume that the ejection by the central engine lasts for a duration $\Delta T$ and consider the energy content of the relativistic outflow is dominated by the kinetic energy. The ejecta shells are then described by a distribution of the Lorentz factor $\Gamma$, width $l$, and mass $m$. Both quantities can vary on a timescale $\Delta t_{\rm c}$ (variability time of the central engine). We model the dynamic of these internal shocks via a multiple-shell model where the successive collisions between shells mimic the propagation of shock waves within the relativistic outflow (see Appendix Sect.\ref{Asec_internal shock modeling}). After modeling internal shock, we can get a series of photon fields, each of which is emitted isotropically in the comoving frame of the thin spherical shell expanding ultra-relativistically with a finite duration.

\subsection{Parameters setting} \label{p set}

\begin{figure*} [htbp]
    \includegraphics[width = 0.5\linewidth]{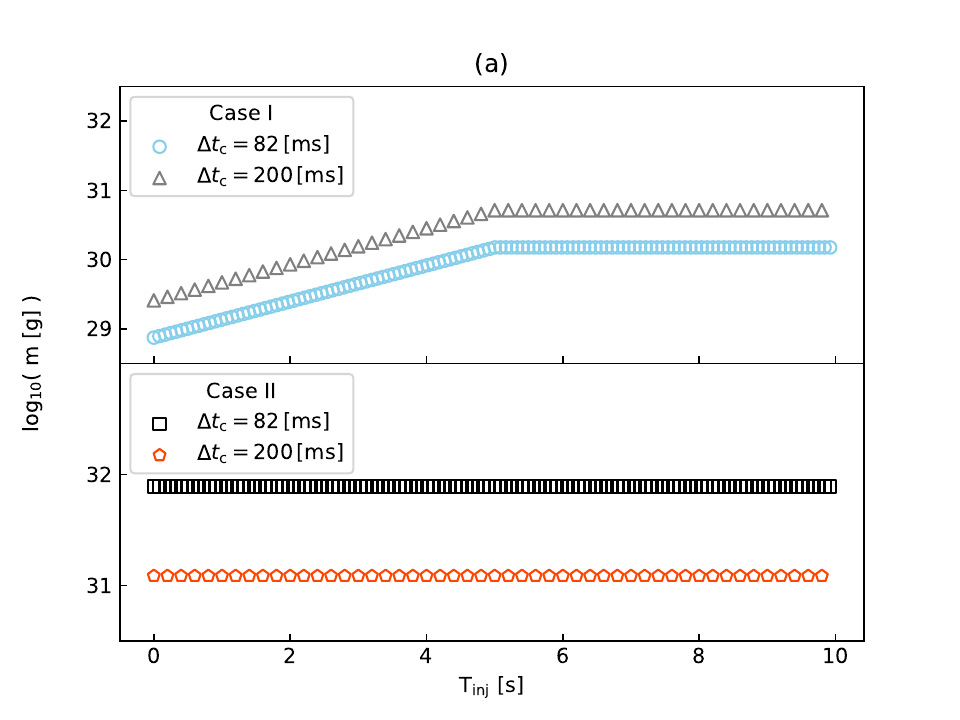}
    \includegraphics[width = 0.5\linewidth]{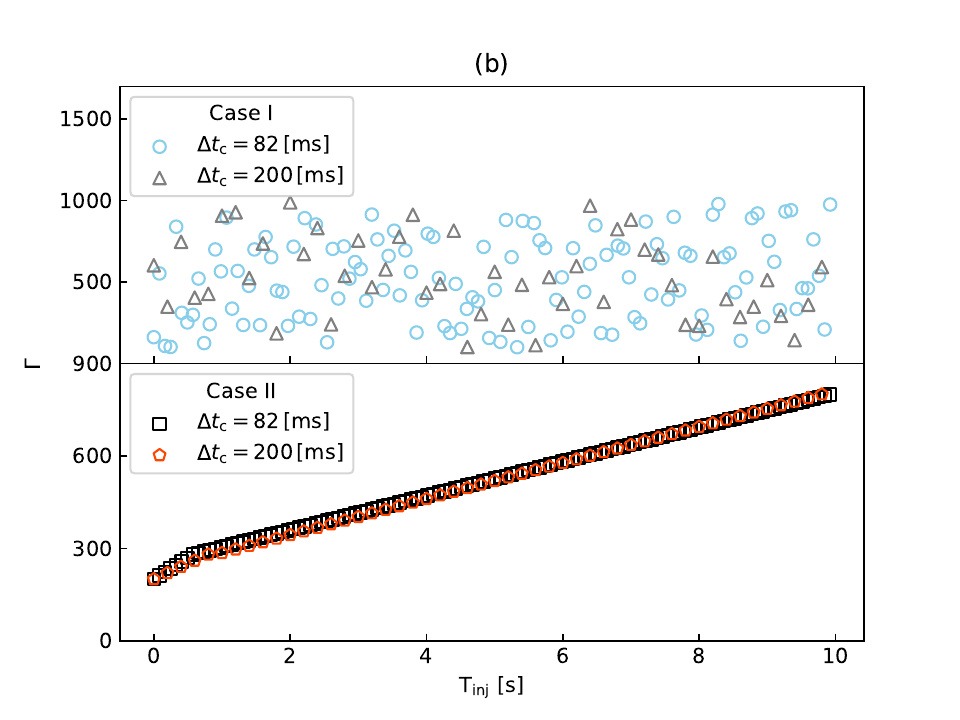}
    \\
    \includegraphics[width = 0.5\linewidth]{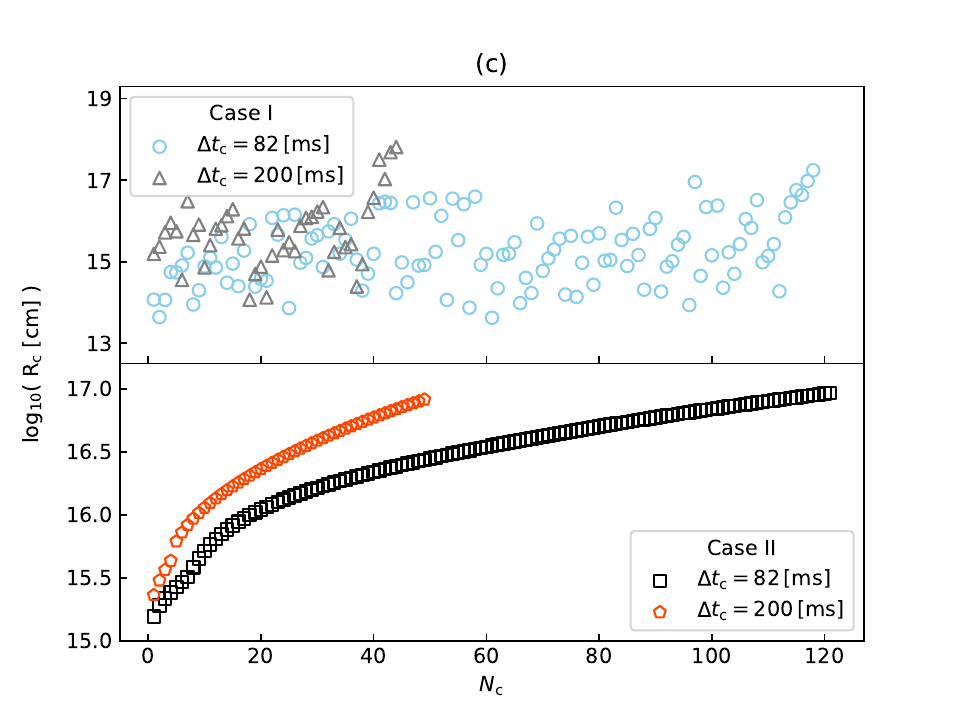}
    \includegraphics[width = 0.5\linewidth]{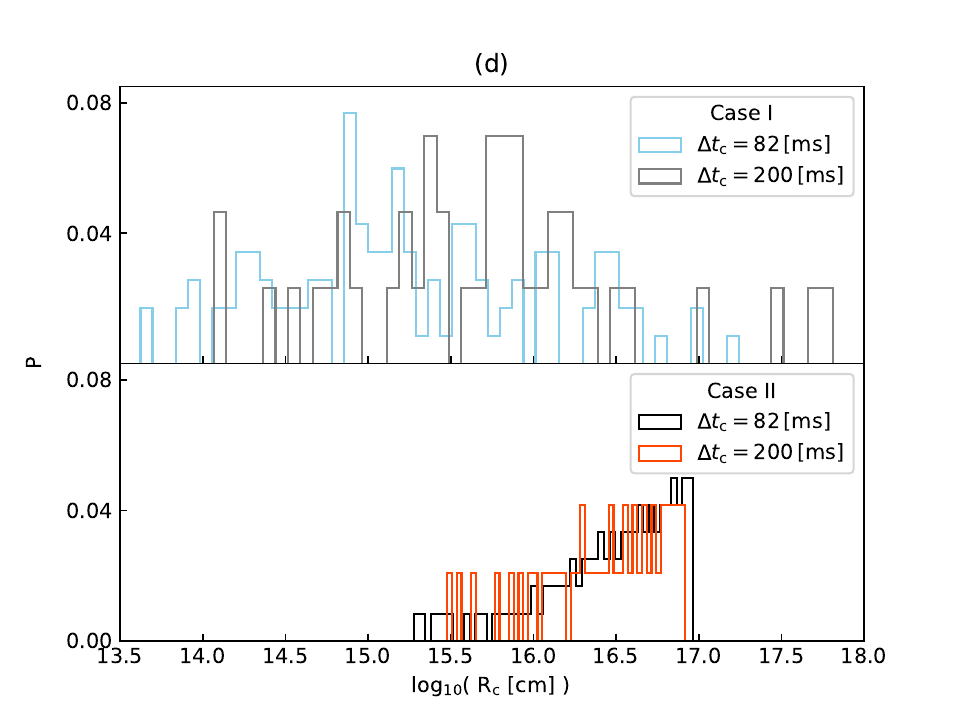}
    \\
    \includegraphics[width = 0.5\linewidth]{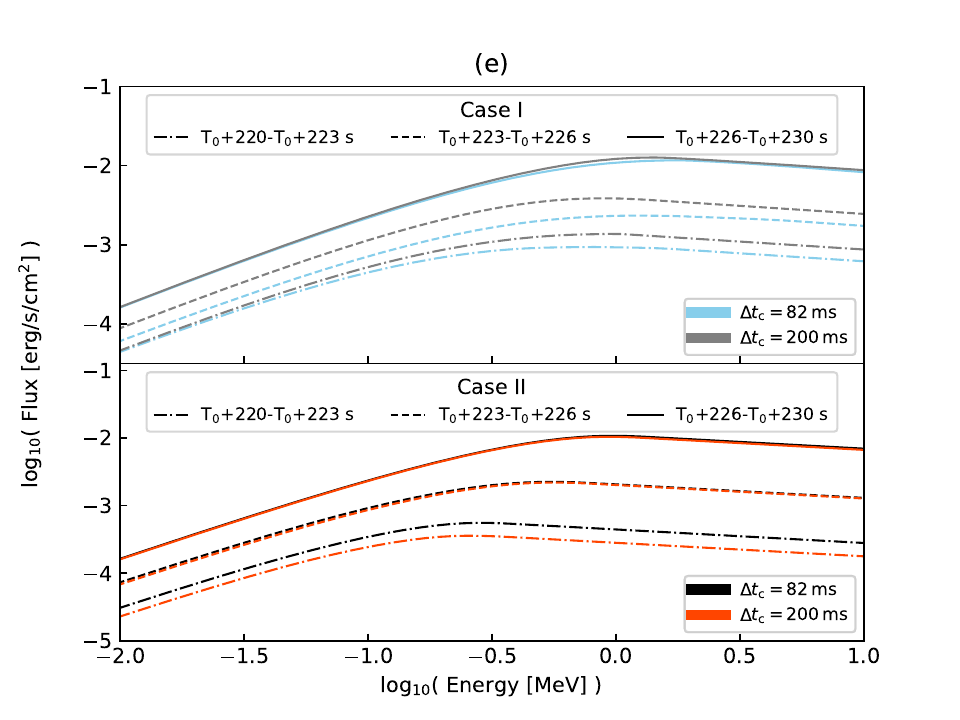}
    \includegraphics[width = 0.5\linewidth]{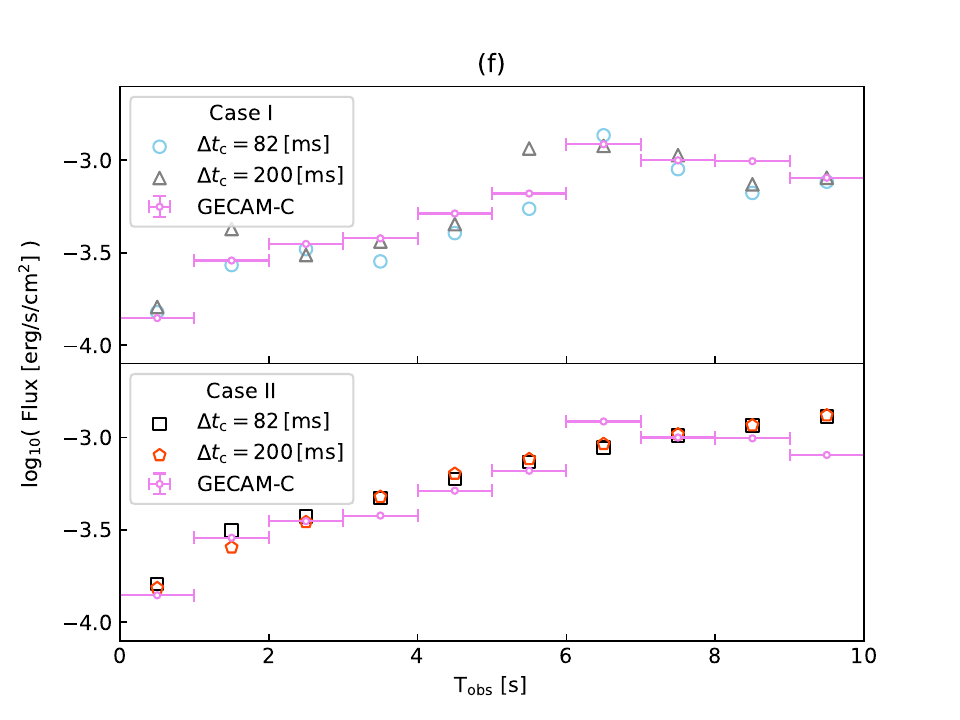}

    \caption{Modeling GRB 221009A in the internal shock scenario through numerical simulation. We use four colors to represent different cases and variability timescales: blue (Case I, $\Delta t_{\rm c} = 82 \, \rm ms$), gray (Case I, $\Delta t_{\rm c} = 200 \, \rm ms$), black (Case II, $\Delta t_{\rm c} = 82 \, \rm ms$) and red (Case II, $\Delta t_{\rm c} = 200 \, \rm ms$). (a) Initial distribution of mass $m$ for the injected shells, with $T_{\rm inj}$ representing the time elapsed after the first shell is injected in the source frame. (b) The initial distribution of Lorentz factors $\Gamma$ for the injected shells. (c)  The collision radius $R_{\rm c}$ for the $N_{\rm c} \rm th$ observed pulse. (d) The distribution of collision radii $R_{\rm c}$, where $P$ is the probability density. (e) The time-average spectra in the observer frame during three time intervals: $T_0+220 \operatorname{-} T_0+223 \, \rm s$ (the dashed-dot line), $T_0+223 \operatorname{-} T_0+226 \, \rm s$ (the dashed line), and $T_0+226 \operatorname{-} T_0+230 \, \rm s$ (the solid line). (f) The synthetic light curve in the $20\operatorname{-}200 \, \rm keV$ range, integrated on equal-arrival time surfaces (EATS) with a time bin of $1\, \rm s$ in the observer frame. The magenta light curve represents the observed one   \citep{GECAM2023}. Light curves with smaller time bins are provided in Fig.\ref{fig_sm_lc} in the Appendix. Here, $T_{\rm obs}$ represents the time elapsed after $T_0+220 \, \rm s$ in the observer frame.}

    \label{fig1}
\end{figure*}

The light curve of GRB 221009A shows a rising phase from about $T_0+220 \, \rm s$ to $T_0+227 \, \rm s$ \citep{GECAM2023, GBM2023}. To reproduce this rising light curve, we explore two scenarios:

Case I: we assume that the mass of the relativistic shells increases with time continuously, with the Lorentz factors $\Gamma$ following the normal distribution in the range from 100 to 1000, centered at an average value of $\Gamma=440$  (see Case I in panel (a) and (b) of Fig.\ref{fig1}).  This averaged  Lorentz factor represents the final Lorentz factor of the merged shell after multiple internal shock collisions, consistent with the inferred Lorentz factor from the TeV afterglow observations of GRB 221009A \citep{LHAASO2023}. 

Case II: in this scenario, we assume that the Lorentz factor of the shells increases with time regularly, ranging from 200 to 800, while the masses of each shell are the same (see Case II in panel (a) and (b) of Fig.\ref{fig1}). This approach results in a much smoother synthetic light curve (see Fig.\ref{fig_sm_lc} in the Appendix), matching  better the observations of GRB 221009A \citep{GECAM2023, GBM2023}.

For the spectrum, we use the Band function  ($\alpha = -0.75$ and $\beta = -2.2$) for all pulses \citep{GECAM2023, GBM2023} (we chose $\alpha = -0.75$ as the observed slope. However, the discussion of the internal shock model's capacity to reproduce this slope is not addressed here). The peak energy $E_{\rm p}$ of the Band function for each  pulse is determined by the $L_{\gamma}\operatorname{-}E_{\rm p}$ relationship: ${L_{\gamma}/(10^{54} \, \rm erg \, s)} = N_0[E_{\rm p}/(1 \, \rm MeV)]^{2}$ \citep{Yonetoku2004, Liang2004}, where $L_{\gamma}$ is the isotropic luminosity and we assume the normalization factor $N_0 = 1$.

We set the initial distance $d = c\Delta t_{\rm c}$ between shells to $5l$. We fix the duration $\Delta T$ of the central engine to $10\,\rm s$. \citet{Liu2023} find that, during the early rising phase ($T_0+210\,{\rm s}$ to $T_0+219\,{\rm s}$) of the main burst emission, at which the detector is not saturated,  the minimum variability time is $\Delta t_{\rm v}=0.082\, {\rm s}$.  \citet{GBM2023} find that the minimum variability time is about 0.2 s during the time right before and after the main burst (see Fig.2 in \cite{GBM2023}). We thus consider two cases of variability time of the central engine: $\Delta t_{\rm c}=82 \, {\rm ms}$ and $\Delta t_{\rm c}=200 \, {\rm ms}$.

\subsection{Simulated light curves and spectra} \label{Simulate}
We then obtain the synthetic light curve
in the 20 - 200 keV range through integrating over equal-arrival time surfaces (EATS), using the parameters outlined in Section \ref{p set}.
The panel (c) of Fig.\ref{fig1} illustrates the collision radius $R_{\rm c}$ for the $N_{\rm c}\rm th$ observed pulse. Notably, in Case II, $R_{\rm c}$ consistently rises with $N_{\rm c}$ because of the regular increase of $\Gamma$. The panel (d) of Fig.\ref{fig1} shows the distribution of the collision radius $R_{\rm c}$. In Case I, with a collision time scale of $\Delta t_{\rm c}=82 \, \rm{ms}$, $R_{\rm c}$ centers at $R_{\rm c}\sim 10^{15} \, \rm{cm}$ with a wide tail. When $\Delta t_{\rm c}=200 \, \rm{ms}$, the center shifts to $R_{\rm c}\sim 2\times 10^{15} \, \rm{cm}$, in agreement with the analytical estimate $R_{\rm c} \sim 2\Gamma_0^2 c \Delta t_{\rm c}$. In Case II, the narrow distribution of $R_{\rm c}$ can also be attributed to the regular distribution of $\Gamma$.

The simulation reproduces some key characteristics of GRB 221009A, as shown in Fig.\ref{fig1}: in panel (e), we present the synthetic GRB spectrum, which roughly matches the observed spectrum (a Band function with $\alpha = -0.75$ and $\beta = -2.2$). The peak energy of the Band function increases with time during the rising phase of the light curve owing to the relationship between $L_{\gamma}$ and $E_{\rm p}$ used in the simulation. Panel (f) of Fig.\ref{fig1} and Fig.\ref{fig_sm_lc} display the MeV emission light curves with different time resolutions. Both light curves exhibit a rising behavior, consistent with the observations of GRB 221009A during the period from $T_0+220 \, \rm s$ to $T_0+230 \, \rm s$ after the trigger.

\section{$\gamma\gamma$ absorption of TeV emission} \label{sec_opacity}
In the internal shock scenario, a series of pulses are generated from collisions. High-energy photons can undergo significant absorption due to the pair-creation process $\gamma\gamma\rightarrow{e^+e^-}$. We define two kinds of absorption according to the origin of the target photons for the annihilation: (1) internal Absorption: this refers to the scenario where a high-energy photon is absorbed by target photons originating from the same pulse as the high-energy photon. (2) external Absorption: in this case, a high-energy photon is absorbed by target photons that originate from other pulses.

An important quantity to describe the degree of absorption is the absorption ratio $f_{\gamma\gamma}$, which is defined as the average ratio of the absorbed flux to that without absorption:
\begin{equation}
    f_{\gamma\gamma}(E_{\rm HE}) = \frac{\sum\limits_{k = 1}^{k_{\rm{tot}}} F_{\nu}(E_{\rm HE}, k)\times e^{-\tau_{\gamma\gamma}(E_{\rm HE}, k)}}{\sum\limits_{k = 1}^{k_{\rm{tot}}} F_{\nu}(E_{\rm HE}, k)},
\end{equation}
where $E_{\rm HE}$ is the energy of a high-energy photon, $F_{\nu}(E_{\rm HE}, k)$ is the flux density of the $k$th pulse, $k_{\rm tot}$ is the number of pulses and $\tau_{\gamma\gamma}(E_{\rm HE}, k)$ is the $\gamma\gamma$ optical depth for the $k$th pulse  (see Eq.\ref{eq_in_abs} and Eq.\ref{eq_ex_abs}).

Before dealing with more complex dynamical configurations within the internal shock framework, we first study a simple single-pulse case and two-pulse case to compare with previous studies and understand the underlying physics.

\subsection{Single-pulse case}  \label{sect single-pulse}
In single-pulse scenario, we assume that photons are emitted isotropically in the comoving frame of a thin spherical shell with a finite duration, which turns on radiation at $R = 10^{15}\, \rm cm$ and turns off at $1.5\times 10^{15}\, \rm cm$ with isotropic luminosity $L_\gamma = 10^{54} \, \rm erg \, s^{-1}$ and Lorentz factor $\Gamma = 440$. In this case, we only need to consider internal absorption within the pulse. The absorption ratio $f_{\gamma\gamma}$ in single-pulse case is calculated by Eq.\ref{eq_in_abs}, which is power-law like (see the black solid line in Fig.\ref{in_ex_result}) instead of exponential. This is in agreement with the result in \cite{Granot2008}, which argues that high-energy photons can escape before the building-up of the target photon field and the time-integrated spectrum is power-law like.

\subsection{Two-pulse case} \label{sect two-pulse}
\begin{figure}
    \centering
    \includegraphics[width = 1.0\linewidth]{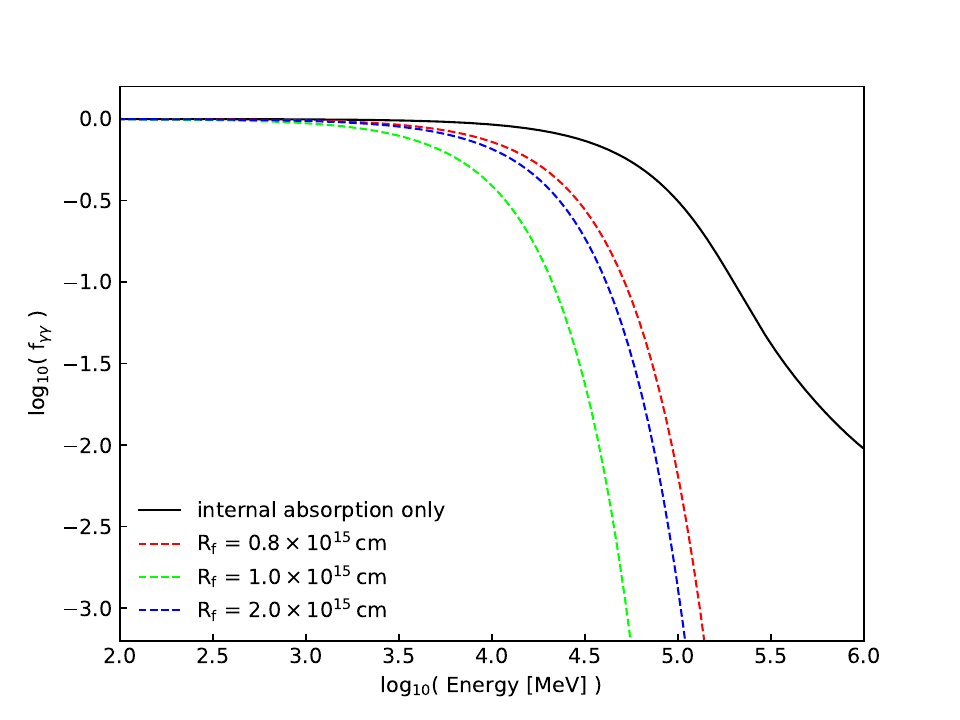}
    \caption{Absorption ratio $f_{\gamma\gamma}$ in single-pulse (the solid line) case and two-pulse case (the dashed lines).}
    \label{in_ex_result}
\end{figure}

For the two-pulse case, we consider two photon fields: the back (inner) photon field absorbed by the front (outer) photon field. Since we only can receive the photons emitted along the line of sight, and the back photons can never catch the front photons emitted on the line of sight, we do not need to consider the absorption of the back photons field to the front photons field. The external absorption of the back photon field can be calculated by Eq.\ref{eq_ex_abs}. We fix the Lorentz factor $\Gamma_{\rm f} = 440$, the total energy $E_{\rm f} = 10^{53} \, \rm erg$ of the front photon field, and the emission radius $R_{\rm b} = 10^{15} \, \rm cm$ of the back photon field. The distance $d_{\rm bf}$ between two photon fields is fixed to $3\times 10^9 \, \rm cm$. Then we vary the emission radius $R_{\rm f}$ of the front photon field and calculate the  $\gamma\gamma$ absorption. The $\gamma\gamma$ opacity is shown in Fig.\ref{in_ex_result} and we find the following results for the external absorption:

(1) While the spectrum of an individual pulse exhibits a power-law distribution, external absorption by other pulses causes the spectrum to become exponential, except for the first received pulse, which remains unaffected by external absorption. This occurs because the front photon field has been established, making the absorption of the subsequent pulse significant.

(2) The strongest absorption of TeV photons in the background photon field occurs when $R_{\rm f}$ is approximately equal to $R_{\rm b}$, as depicted in Fig.\ref{in_ex_result}. This happens because: (i) when $R_{\rm f} < R_{\rm b}$, the absorption region extends beyond $R_{\rm b}$. In such cases, a small $R_{\rm f}$ leads to significant anisotropy in the absorption region, resulting in a low $\gamma\gamma$ opacity. (ii) conversely, if $R_{\rm f} > R_{\rm b}$, the absorption region surpasses $R_{\rm f}$. In this situation, a large $R_{\rm f}$ results in a lower energy density within the absorption region, leading to reduced $\gamma\gamma$ opacity.

\subsection{The $\gamma\gamma$ absorption in internal shocks}
\begin{figure*} [htbp]
    \includegraphics[width = 0.5\linewidth]{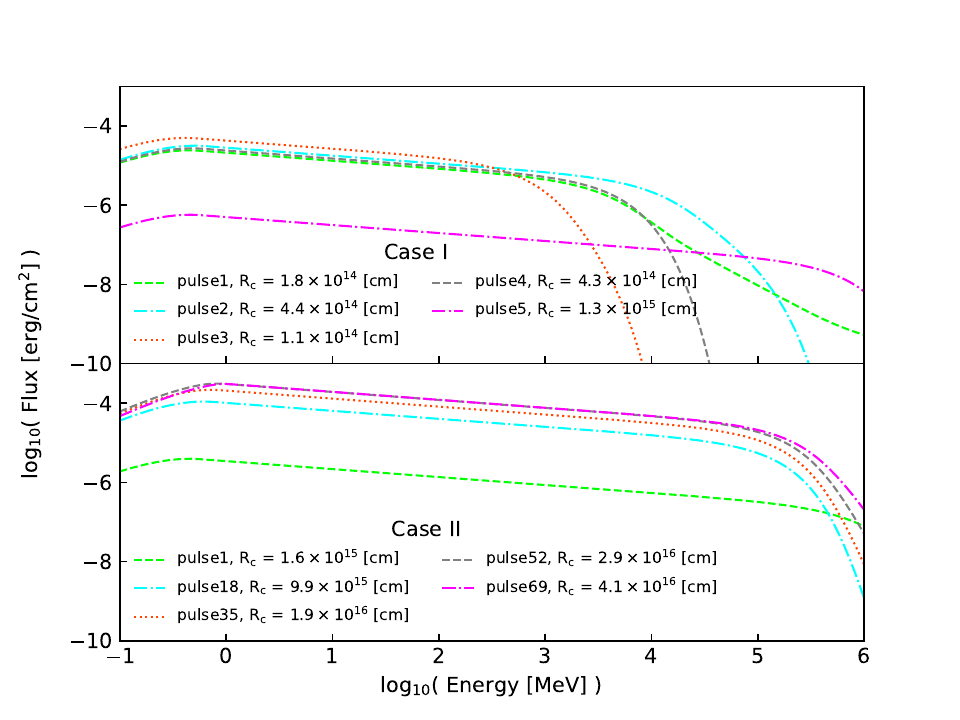}
    \includegraphics[width = 0.5\linewidth]{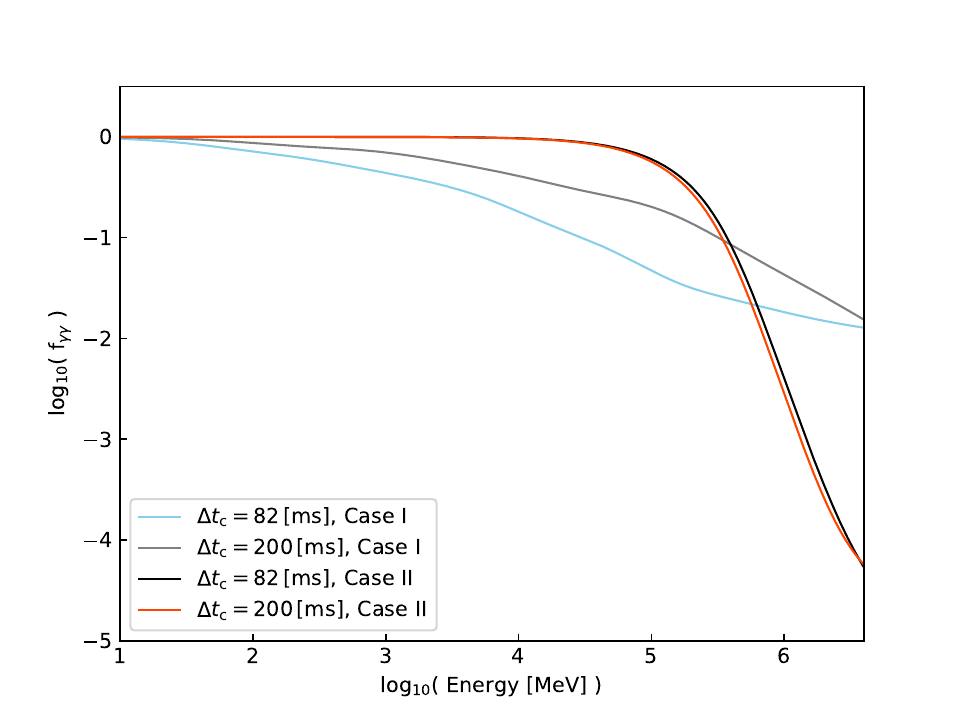}\\
    
    \caption{Left Panel: The spectra of five pulses in the simulation for both Case I and Case II (refer to Sect.\ref{p set}), using $\Delta t_{\rm c}=82\,\rm ms$. In Case I, we choose the first five observed pulses; In Case II, we choose 5 pulses with intervals of 17 pulses to see the evolution of cut-off energies from different pulses. Right Panel: The time-integrated absorption ratio $f_{\gamma\gamma}$ for the simulated GRB, which reproduces the key characteristics of GRB 221009A.}
    \label{abs_result}
\end{figure*}

In this section, the model is applied to the dynamic evolution anticipated within the internal shock framework, where the entire prompt gamma-ray emission is considered as a collection of multiple sub-pulses (refer to Sect.\ref{Simulate}).

\begin{figure}
    \centering
    \includegraphics[width = 1.0\columnwidth, angle = 0, scale = 1.0]{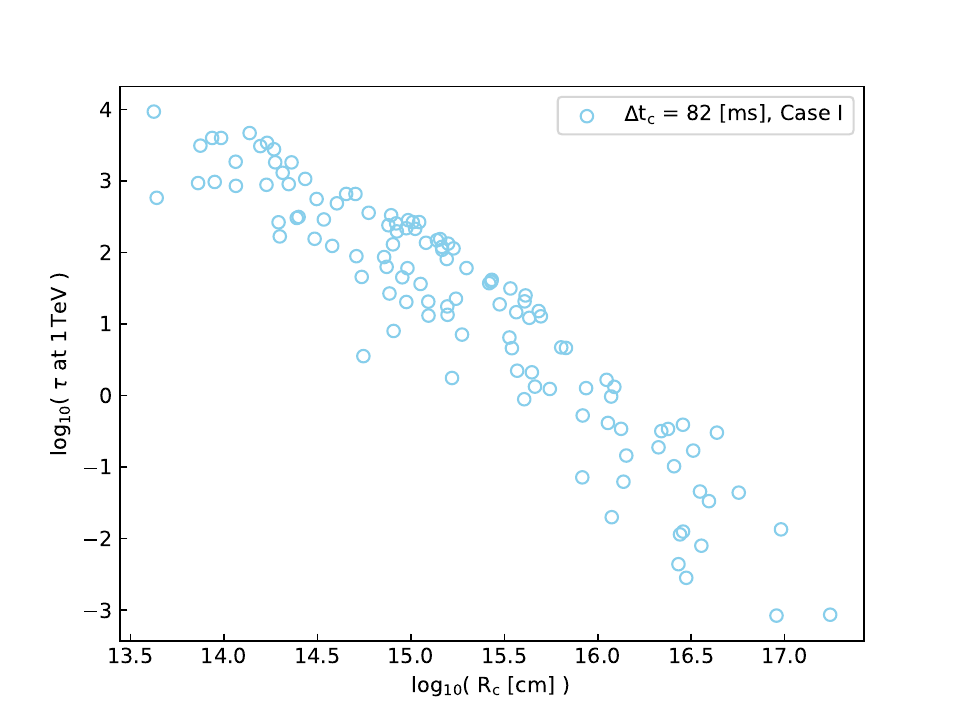}
    \caption{The $\gamma\gamma$ absorption optical depth $\tau$ at $1\,\rm TeV$ of the pulses generated from internal shock simulations as a function of the emission radius of pulses in Case I (see Sect.\ref{p set}) with $\Delta t_{\rm c} = 82 \, \rm ms$.}
    \label{fig_tau_R}
\end{figure}

Case I: in the left panel of Fig.\ref{abs_result}, we present the spectra of the first five observed pulses from the simulation (as discussed in Sect.\ref{p set}) with a collision time scale of $\Delta t_{\rm c} = 82 \, \rm ms$. The initial pulse (pulse1 in Fig.\ref{abs_result}) exhibits a power-law-like spectrum as it does not experience external absorption. However, all subsequent pulses display an exponential cutoff due to external absorption. This result agrees with the results in Sect.\ref{sect single-pulse} and Sect.\ref{sect two-pulse}.
The time-integrated spectrum of all pulses becomes notably flatter and resembles a power-law due to the superposition of the spectra with different cutoff energies ($E_{\rm cut}$) where $\tau_{\gamma\gamma}(E_{\rm cut}) = 1$. $E_{\rm cut}$ primarily depends on two factors: the energy density of the local photon field and the degree of anisotropy of the photon field. At larger radii, where the photon energy density is low and the photon field anisotropy is high, the $\gamma\gamma$ opacity decreases (as seen in Fig.\ref{fig_tau_R}), resulting in a larger $E_{\rm cut}$. The result that the superposition of spectra with different cutoff energies leads to a power-law spectrum is consistent with the finding in \cite{Aoi2010}, although no external absorption is considered in \cite{Aoi2010}.

Case II: we also present five pulses from the simulation for this case in the left panel of Fig.\ref{abs_result}. Similar to Case I, the first observed pulse exhibits a power-law-like spectrum because it does not suffer external absorption. However, in Case II, the cutoff energies across different pulses are similar, in contrast to Case I. This similarity arises from the regular distribution of the Lorentz factors ($\Gamma$), which results in a tight distribution of collision radii $R_{\rm c}$ (as seen in panels (c) and (d) in Fig.\ref{fig1}). Most of the pulses occur in regions with similar photon energy density and photon field anisotropy due to the tight distribution of $R_{\rm c}$, leading to similar cutoff energies. The superposition of multi-pulse spectra still forms a power-law spectrum because of two factors: (1) the large collision radii lead to a large cutoff energy which is close to the observation window; (2) photons from the first pulse can escape before the photon field fully develops. However, in this case, the power law is steeper than that in Case I, with $f_{\gamma\gamma}\propto E^{-2.5}$ (see the right panel of Fig.\ref{abs_result}).

We display the absorption ratio $f_{\gamma\gamma}$ as a function of photon energy in the right panel of Fig.\ref{abs_result}. In Case I, a smaller time variability leads to a larger absorption. This is because a smaller variability time results in smaller collision radii and a more intense photon energy density. Conversely, in Case II, the variability time does not significantly impact the absorption ratio. This is because different variability time primarily affects the number of injected shells while the collision radii remain the same (see panel (d) of Fig.\ref{fig1}). This leads to similar photon energy densities and, consequently, similar levels of absorption.

\section{Constraints on the jet composition in the internal shock scenario} \label{sec_constrain}
The obtained absorption ratio $f_{\gamma\gamma}$ is insufficient to explain the flux ratio $F_{\rm TeV}/F_{\rm MeV}\le   2\times10^{-5}$ if the SSC emission component (i.e., TeV emission) has a comparable flux to that of the synchrotron component (i.e., MeV emission). This suggests a low ratio between the SSC and synchrotron outputs. Below we discuss its implication for the magnetic field intensity in the emitting region.

We assume that in GRB internal shocks, fractions of $\epsilon_{B}$ and $\epsilon_{e}$ of the shock internal energy are converted into the
energy in the magnetic field and electrons, respectively. It is
usually assumed that the electrons are rapidly cooling, so the
energy density in gamma-ray emission $U_\gamma$ is equal to the
electron energy density $U_e$. The magnetic field is given by:

\begin{equation}
\frac{B^2}{8\pi}=\left(\frac{ \epsilon_{B}}{ \epsilon_{e}}\right)U_\gamma=\left(\frac{ \epsilon_{B}}{ \epsilon_{e}}\right)\frac{L_\gamma}{4\pi
R_{\rm in}^2 c \Gamma_0^2},
\end{equation}
where $R_{\rm in}$ is the radius of the internal shock dissipation, $L_\gamma$ is the luminosity
in gamma-ray emission and $\Gamma_0$ is the bulk Lorentz factor. In the synchrotron model for the prompt MeV
emission, by use of $h\nu_{
m}=\phi_{\nu}\frac{3hqB}{4\pi m_e c}\gamma_m^2\Gamma$,  one can
derive the Lorentz factor of electrons that radiate at the GRB
peak energy $h\nu_m$~\citep{Wang2009},

\begin{equation}
\label{eq:gammam}
\begin{array}{ll}
\gamma_m &=\left(\frac{4\pi m_e c}{3\phi_\nu hq}\right)^{1/2}
\left(\frac{ \epsilon_{}}{ \epsilon_{B}}\right)^{1/4}\left(\frac{2 L_\gamma}{R^2
c}\right)^{-1/4} \varepsilon_p^{1/2}\\
&=1.5\times 10^3 \left(\frac{ \epsilon_{e}}{ \epsilon_{B}}\right)^{1/4}  \left(\frac{L_{\gamma}}
{3\times10^{53}}\right)^{-1/4} R_{\rm in,15}^{1/2} \left(\frac{h\nu_m}{1\, {\rm
MeV}}\right)^{1/2},
\end{array}
\end{equation}
where $\phi_\nu\simeq0.5$ is the coefficient defined in \cite{Wijers} and $q$ is the electron charge.

The peak energy of the prompt emission during the initial phase of GRB 221009A is ${\sim 1}$~MeV \citep{GCN-KW}, so we take $h\nu_m\sim 1\, {\rm MeV}$. The KN limit introduces a new critical Lorentz factor $\hat\gamma_m$, defined as $\hat\gamma_m\equiv \Gamma m_e c^2/h\nu_m$, corresponding to the critical electrons that up-scattering photons at energy $h\nu_m$ in the Thompson scattering regime \citep{Nakar2009}. We obtain $\hat\gamma_m=220 (\Gamma_0/440) (h\nu_m/ 1\, {\rm MeV})^{-1} $.  Unless $ \epsilon_{e}\le 10^{-4} \epsilon_{B}$, which is unreasonable in
terms of the burst energetics, we have $\gamma_m>\hat\gamma_m$, so the IC scattering between
$\gamma_m$-electrons and the bulk of the gamma-ray emission should
be in the Klein-Nishina (KN) regime. In the KN regime,  the  SSC spectrum is given by \citep{Nakar2009}:

\begin{equation}
F_{\nu}= \left\{
\begin{array}{lll}
\nu^{-\frac{1}{2}},  \,\,\,\, 2\gamma_c^2\nu_c<\nu<2\nu_m\gamma_m{\hat\gamma_m} \\
\nu^{-p+\frac{1}{2}}, \,\,\,\, \nu> 2\nu_m\gamma_m{\hat\gamma_m}.  \\
\end{array}
\right.
\end{equation}
The SSC emission flux peaks at
\begin{equation}
\begin{array}{ll}
\varepsilon_p^{\rm IC}&=2\nu_m\gamma_m{\hat\gamma_m}=2\Gamma_0\gamma_m m_e c^2\\
&=0.7\,{\rm TeV}\, \left(\frac{L_{\gamma}}
{3\times10^{53}}\right)^{-1/4} \left(\frac{\Gamma_0}{440}\right) \left(\frac{ \epsilon_{e}}{ \epsilon_{B}}\right)^{1/4} R_{\rm in,15}^{1/2} \left(\frac{\varepsilon_p}{1\, {\rm
MeV}}\right)^{1/2},
\end{array}
\end{equation}
where we have used Eq.\ref{eq:gammam}. Thus the SSC spectral peak is expected to be located within or close to the observed energy range of LHAASO.  Defining $f_{\rm spec}$ as a suppression factor due to SSC peak deviation from the LHAASO energy range,  we expect $f_{\rm spec}\ge 0.3$ in a conservative way.

In the case of $\gamma_m>\hat\gamma_m$, the SSC to synchrotron energy output is well approximated by \citep{Nakar2009}:
\begin{equation}
\label{Y}
Y=\frac{L_{\rm SSC}}{L_{\rm syn}}=\frac{ \epsilon_{e}}{ \epsilon_{B}}f_{\rm KN}\sim (p-2)\frac{ \epsilon_{e}}{ \epsilon_{B}}\left(\frac{\gamma_m}{\hat\gamma_m}\right)^{-1/2},
\end{equation}
where the suppression due to the KN effect is $f_{\rm KN}=(p-2)\left(\frac{\gamma_m}{\hat\gamma_m}\right)^{-1/2}\sim 0.1\left( \frac{\epsilon_e}{\epsilon_B} \right)^{-1/8}$ for typical parameter values.

During $T_0+220\,{\rm s}$ to $T_0+230\,{\rm s}$,  the flux ratio between TeV flux and the MeV flux is $\bar{R}\equiv F_{\rm TeV}/F_{\rm MeV}\le 2\times 10^{-5}$.
Considering the $\gamma\gamma$ absorption, the SSC to synchrotron luminosity ratio should be $Y\le \bar{R}/(f_{\rm spec}f_{\gamma\gamma})$. From Eq.\ref{Y}, we obtain roughly
\begin{equation}
\left\{
\begin{array}{l}
    \epsilon_{B}\ga 45 \epsilon_{e} , \,\rm case\, I \, (\Delta t_{\rm c} = 82\, \rm ms)\\
    \epsilon_{B}\ga 118 \epsilon_{e} , \,\rm case\, I \, (\Delta t_{\rm c} = 200\, \rm ms)\\
    \epsilon_{B}\ga 12 \epsilon_{e} , \,\rm case\, II \, (\Delta t_{\rm c} = 82\, \rm ms)\\
    \epsilon_{B}\ga 13 \epsilon_{e} , \,\rm case\, II \, (\Delta t_{\rm c} = 200\, \rm ms)
\end{array}
\right.
\end{equation}
for typical value of $p=2.2\operatorname{-}2.5$. This means that the magnetic field energy density is much larger than the electron energy density, so the SSC flux can be greatly suppressed. This is inconsistent with the usual assumption that energy equipartition between the ion, electron, and magnetic energy is obtained at the internal shocks \citep{Murphy2010}. $ \epsilon_{B}\gg \epsilon_{e}$  implies a small $\epsilon_e$ for GRB 221009A, which leads to a small radiation efficiency for internal shock emission. Therefore, the prompt TeV limit imposed by LHAASO observations seems to challenge the standard internal shock model for GRB 221009A.

\section{The magnetic field dissipation model} \label{mangetic dissipation}

In the above, we have assumed a matter-dominated composition, where the internal shock scenario applies.  $\epsilon_{B}\gg \epsilon_{e}$ points to a magnetic-field dominated composition of the jet. If the GRB ejecta carries a dynamically important magnetic field
component, i.e., $\sigma>1$,  the GRB radiation is thought to be powered by the
dissipation of the magnetic field energy in the ejecta \citep{Meszaros1997, Lyutikov2003, Narayan2009}.

In the Internal-Collision-Induced Magnetic Reconnection and Turbulence (ICMART) model \citep{zhang2011}, magnetically-dominated winds are injected from the central engine intermittently. At first, the field lines are ordered and have the same orientation, so the reconnection is suppressed. However, the collisions trigger an “avalanche” of magnetic reconnection/turbulence events, causing efficient radiation. So the ICMART model needs multiple collisions to dissipate magnetic energy and cause radiation, which tends to occur at larger radii. In the ICMART scenario, we consider a single main pulse emitted with an observed duration of about $5\,\rm s$). We assume that the pulse turns on radiation at $\sim R_{\rm ICMART} = 10^{16} \, \rm cm$ or $10^{17} \, \rm cm$, respectively, and other parameters are the same as those in the back pulse of Fig.\ref{in_ex_result}. In this case, we find the TeV absorption ratio decreases significantly, with a value of $f_{\gamma\gamma} = 0.34$ and $0.82$ for $R_{\rm ICMART} = 10^{16}$ and $10^{17} \, \rm cm$, respectively, at 1 TeV.  

The suppression ratio due to the KN effect is  $f_{\rm KN}\propto R^{-1/4}$ according to Eq.\ref{eq:gammam} and $f_{\rm KN}\propto \gamma_m^{-1/2}$. We obtain $f_{\rm KN} \simeq 0.06$ and $0.03$ for $R_{\rm ICMART}=10^{16}\, \rm cm$ and $10^{17}\, \rm cm$, respectively. In the ICMART scenario, the gamma-ray luminosity is $L_\gamma=L_w \eta \epsilon_e$, where   $L_w$ is the total isotropic luminosity of the wind and $\eta$ is the energy dissipation efficiency \citep{zhang2011}. Then the energy density of photons in the comoving-frame of the jet is $U'_{ph}=\frac{L_\gamma}{4\pi R^2 \Gamma^2 c}$ and the energy density of the magnetic field is $U'_B=\frac{L_w}{4\pi \Gamma^2 R^2 c} \frac{\sigma}{1+\sigma}$. Then we have:
\begin{equation} \label{Y_ICMART}
    \frac{L_{\rm SSC}}{L_{\rm syn}} = \frac{U'_{ph}}{U'_B}f_{\rm KN} = \frac{L_w\eta \epsilon_e}{L_w\left( \frac{\sigma}{1+\sigma} \right)} f_{\rm KN}.
\end{equation}
Using $\bar{R}\equiv F_{\rm TeV}/F_{\rm MeV}\le 2\times 10^{-5}$ and assuming $\epsilon_e = 0.1$, we get $\eta\le 0.033$ for $R_{\rm ICMART} = 10^{16} \, \rm cm$  and $\eta\le 0.027$ for  $R_{\rm ICMART} = 10^{17} \, \rm cm$  when $\sigma\gg 1$. 

Note that the above estimate of the efficiency $\eta$ applies only to $T_0+220\operatorname{-}T_0+230\, \rm s$, during which about $1/3$ of the whole prompt radiation energy is radiated according to prompt emission observations \citep{GECAM2023, GBM2023}. The efficiency at a later time could be higher as $\sigma$ decreases. In addition, the above calculation has assumed that the property of the emitting region of ICMART is the same as that in the internal shock, which is not the case. In the ICMART model, the observed flux is the superposition of the many mini-jets with different orientations.  Each reconnection event is a fundamental mini-jet in the ICMART model, and the direction of the mini-jets can be isotropic \citep{Zhang2014, Shao2022}. This would lead to a larger interaction angle for high-energy photons and a stronger $\gamma\gamma$ absorption. Therefore the absorption factor $f_{\gamma\gamma}$ could be lower than the above estimate and the inferred radiation efficiency will increase correspondingly.   Detailed calculation of the absorption ratio including mini-jet structure is beyond the scope of this work.

\section{Conclusions and Discussions} \label{summary}
\cite{LHAASO2023} estimated the $\gamma\gamma$ optical depth in the prompt emission of GRB 221009A to be $\tau_{\gamma\gamma}\sim 190$ assuming an isotropic radiation field in the single-zone model.  
In this work, we find that although the optical depth is large, it does not lead to exponential attenuation of TeV photons in the framework of the internal shock scenario. We find that the superposition of spectra of different pulses with different cut-off energy results in a power-law-like spectrum for the time-integrated emission (see Fig.\ref{abs_result}). The absorption ratio is only about $\sim 10^{-1.5} \operatorname{-} 10^{-2.5}$ for TeV photons, assuming a bulk Lorentz factor of $440$ and the variability time in the range of $82\operatorname{-}200 {\, \rm ms}$.  This absorption ratio cannot explain the low  TeV flux limit of GRB 221009A imposed by the LHAASO observations and requires a low ratio between the SSC and synchrotron emission outputs. Then we find that  $\epsilon_{B}/\epsilon_{e}\ga 10\operatorname{-} 100$ for GRB 221009A.

The inference  $\epsilon_{B}\gg \epsilon_{e}$ is inconsistent with the usual assumption that energy equipartition between the ion, electron, and magnetic energy is obtained in internal shocks \citep{Murphy2010}. $\epsilon_{B}\gg \epsilon_{e}$  also leads to a small radiation efficiency for internal shock emission. On the other hand, $\epsilon_{B}\gg \epsilon_{e}$ seems to point to a Poynting-flux dominated outflow in which the radiation is powered by the magnetic dissipation. We discussed the possibility of interpreting the prompt emission of GRB 221009A with the ICMART model. The $\gamma\gamma$ absorption could be larger due to the configuration of mini-jets, although a detailed calculation is needed to verify this. If this is the case, then a mild magnetization with $\sigma \sim {\rm a \, few}$ would be sufficient to explain the TeV limit and also lead to a high radiation efficiency.

\appendix 
\setcounter{equation}{0}
\renewcommand\theequation{A\arabic{equation}}

\section{method} \label{Asec_appendix}
\subsection{Internal shock modeling} \label{Asec_internal shock modeling}
We model the internal shock process following the approach in \cite{Kobayashi1997}. At $t = 0$, all the shells are at initial positions $R_i$ and characterized by the following parameters: width $l_i$, mass $m_i$, and the Lorentz factor $\Gamma_i$. So the distance $d_i$ between shell $i$ and shell $i+1$ is $R_i-R_{i+1}-l_{i+1}$. The internal collisions keep occurring until all the shells merge into one shell or the velocities of all inner shells are slower than that of outer shells. We note that the relative value of the mass of shells determines the dynamic evolution rather than the absolute value. The absolute value of the mass of shells only determines the total energy of radiation. For the $j$th collision, we can calculate and record the following physical quantities in the source frame: collision radius $r_j$, collision time $t_j$, duration time $\delta t_j$ of emission and energy $E_j$ of emission and Lorentz factor of the merged shell $\Gamma_{{\rm m},j}$. By recording these quantities, we can calculate both internal and external absorption. 

\subsection{Internal absorption} \label{Asec:_internal absorption}
As mentioned above, photons are emitted isotropically in the comoving frame with finite duration by a thin spherical shell. We use numerical method to calculate $\gamma \gamma$ opacity as in \cite{Hascoet}:
\begin{equation}
\begin{array}{ll} \label{eq_abs}
    \tau_{\gamma\gamma}(E_{\rm{HE}}, j, k) = \sum\limits_{k'=1}^k \frac{\sigma_{\rm{T}} {\epsilon}_{{\rm{rad}},jk'}}{4\pi R_{jk'}^2\Gamma_{jk'}E_{{\rm{p}},jk'}'} \\
    \times \int \, d\ell \, \mathcal{F} [\ell; E_{\rm{HE}}, R_{jk}, t_{jk}; R_{jk'}, t_{jk'}, E_{{\rm{p}}, jk'}', {\mathcal{B}_{jk'}}].
\end{array}
\end{equation}
Here subscript $j$ indicates the quantities of the $j$th collision. We divide the duration of the $j$th emission into a series of time grids so that in every grid the emission can be approximated as a “flash” instead of continuous radiation. $t_{jk}$, $R_{jk}$, $\Gamma_{jk}$, $\epsilon_{{\rm{rad}}, jk}$ and $E'_{{\rm{p}}, jk}$ $(k=1,\cdots, k_{\rm{max}})$ are the emission time, emission radius, Lorentz factor, total radiation energy and spectra peak energy of the $k$th flash in the $j$th emission. The emission turns on at $t_{j1} = t_j$ and turns off at $t_{jk_{\rm{max}}} = t_j+\delta t_j$, where $t_j$ and $\delta t_j$ is the collision time and duration of emission of $j$th collision. We assume $\Gamma_{jk}$, $E'_{{\rm{p}},jk}$, normalized spectra $\mathcal{B}_{jk'}$ ($\int_0^{\infty} \mathcal{B}(x)\,dx = 1$, $x=E_{\rm HE}/E_{\rm p}$) and luminosity $L_{{\rm{m}},j}$ ($L_{{\rm{m}},j} = \sum \limits_{k'=1}^{k_{\rm max}} \epsilon _{{\rm rad},jk'}/{\delta t_j}$) are constant during the shell spreading in single collision event. All quantities in Eq.\ref{eq_abs} are in source frame besides $E'_{{\rm{p}}, jk}$, which is in comoving frame, and normalized spectra $\mathcal{B}$ are the same in source frame and comoving frame.

The internal $\gamma \gamma$ opacity of the $j$th pulse can be written as:


\begin{equation} \label{eq_in_abs}
    \tau_{\gamma \gamma,\rm in}(E_{\rm{HE}}, j) = -ln\left[ f_{\gamma\gamma,\rm in}(E_{\rm HE}) \right] = -ln \left[ \frac{\sum\limits_{k = 1}^{k_{\rm{max}}} F_{\nu}(E_{\rm HE}, j, k)\times e^{-\tau_{\gamma\gamma}(E_{\rm HE}, j, k)}}{\sum\limits_{k = 1}^{k_{\rm{max}}} F_{\nu}(E_{\rm HE}, j, k)} \right].
\end{equation}

All the quantities requiring astrophysical input are $R_j$, $\delta t_j$, $L_{{\rm{m}},j}$, $\Gamma_{{\rm m},j}$ and photon index of spectra to calculate $\tau_{\gamma \gamma,\rm in}$. $R_j$, $\delta t_j$, $L_{{\rm{m}},j}$ and $\Gamma_{{\rm m},j}$ can be calculated from internal shock modeling. The Photon index should be determined by observation and we assume the photon index of the spectrum is the same for all collisions.

\subsection{External absorption} \label{Asec: external absorption}
In order to calculate $\gamma \gamma$ opacity between different shells and reduce computing time, we make an approximation that every emission of collision is a flash. If the shell's distance $d_i$ is larger than the shell's width $l_i$, then the pulse width $\delta t_j$ is determined by angular spreading time instead of hydrodynamic time \citep{Kobayashi1997}. So flash approximation is effective since we assume $d_i > l_i$ in the internal shock modeling. By using this approximation, we can calculate the external $\gamma \gamma$ opacity of the $j$th pulse:

\begin{equation} \label{eq_ex_abs}
\begin{array}{ll}
    \tau_{\gamma\gamma, \rm ex}(E_{\rm{HE}}, j) = \sum\limits_{j'} \frac{\sigma_{\rm{T}} {\epsilon}_{{\rm{rad}},j'}}{4\pi R_{j'}^2\Gamma_{j'}E_{{\rm{p}},j'}'} \\
    \times \int \, d\ell \, \mathcal{F} [\ell; E_{\rm{HE}}, R_{j}, t_{j}; R_{j'}, t_{j'}, E_{{\rm{p}}, j'}', {\mathcal{B}_{j'}}].
\end{array}
\end{equation}

All the pulses outside of the $j$th pulse should be included in Eq.\ref{eq_ex_abs} since only photons from outside of the $j$th pulse can catch the photons from the $j$th pulse in the external scenario as that mentioned in Sect.\ref{sect two-pulse}. The quantities requiring astrophysical input in Eq.\ref{eq_ex_abs} are collision radius $R_j$, collision time $t_j$, Lorentz factor of merged shell $\Gamma_{{\rm m},j}$, energy of emission $\epsilon_{{\rm rad}, j}$ and spectrum $\mathcal{B}_{j}$, all have been calculated in the internal shock modeling besides $\mathcal{B}_{j}$, which has been determined by observation as mentioned above.

\section{The synthetic light curve of GRB 221009A} \label{Bsec_appendix}

\begin{figure*} [htbp]
    \includegraphics[width = 1.0\linewidth]{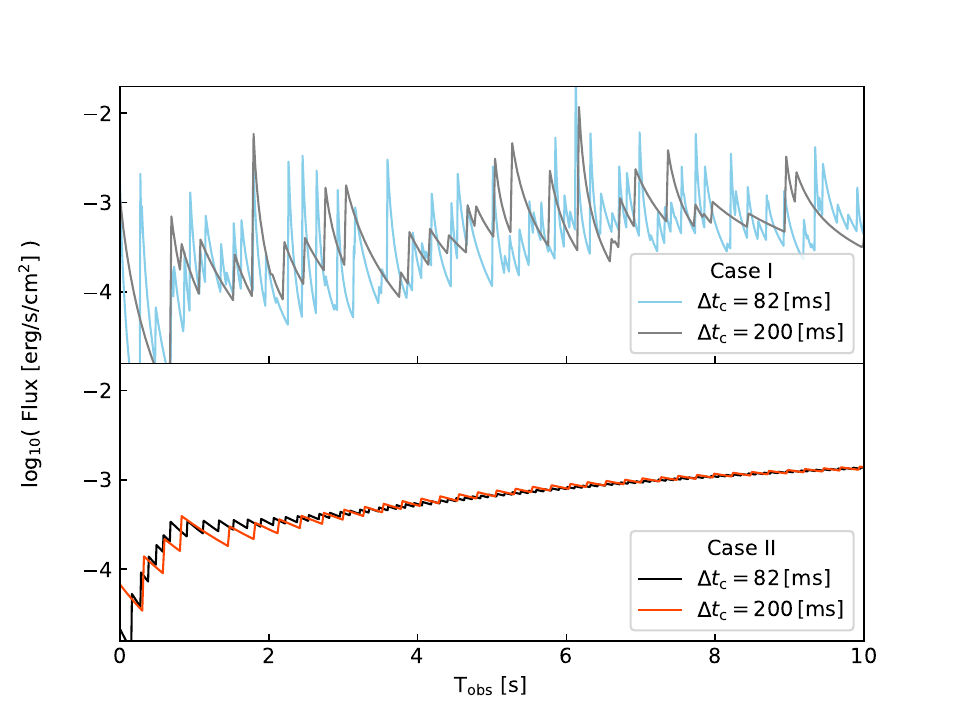}
    \caption{The simulated light curve in $20\operatorname{-}200 \, \rm keV$ of GRB 221009A, assuming a time resolution $\ll 82 \, \rm ms$.}
    \label{fig_sm_lc}
\end{figure*}

\clearpage

\bibliography{ref}{}
\bibliographystyle{aasjournal}

\end{document}